# Specular Reflection Leads to Maximum Reduction in Thermal Conductivity

Martin Maldovan

*School of Chemical & Biomolecular Engineering and School of Physics*
*Georgia Institute of Technology,*
*311 Ferst Drive, Atlanta, GA 30332 USA*

In contrast to established work that use diffuse surface scattering as the mechanism to lower the thermal conductivities, we show that the largest reduction of heat conduction in thin films is achieved via specular scattering. Our results create a new paradigm for heat conduction manipulation since smooth surfaces – in contrast to rough surfaces – are shown to be more effective on suppressing heat conduction.

In the past decades it has been demonstrated that the thermal conductivity is not a fixed physical property but it can be controlled by modifying the properties of thermal phonons [1]. A large number of experiments have been reported where phonon mean free paths are reduced by orders of magnitude [2-6]. More recently, there have also been experiments in which thermal phonon group velocities have been modified via wave effects [7,8]. The ability to manipulate thermal transport properties is of interest in a wide variety of fields ranging from thermoelectric to electronic devices [9,10]. Ideally, it would be desirable to achieve both low and high thermal conductivities since access to low thermal conductivities allows for more efficient thermoelectrics while high thermal conductivities are critically needed for dissipation in electronics and optoelectronics.

A fundamental question that remains is whether diffuse or specular surface scattering of phonons can enhance or reduce heat conduction. To date, it is well established that diffuse surface scattering causes a reduction of heat conduction at small length scales [1]. In recent years, diffuse surface scattering has been widely used to achieve low thermal conductivities in nanowires, superlattices, nanoporous, and nanocomposite structures [2,3,6,11-15]. In addition, it has been demonstrated that coherent interference of thermal phonons reduces heat conduction in nanoporous and superlattice structures by decreasing phonon group velocities [7,8]. Contrarily to the large amount of work on reducing heat conduction, approaches to enhance heat conduction via nanostructuring have been limited. It has recently been shown that specular phonon scattering can enhance heat conduction by enabling phonon spectral coupling [16]. Importantly, based on several studies on nanoscale thermal phonon transport, the role of diffuse surface scattering has been to provide access to a broad spectrum of nanomaterials with decreased thermal conductivities [2-6,9-15].

In contrast to established work that consider diffuse surface scattering as the mechanism to lower the thermal conductivity, here we show that the greatest possible reduction of heat conduction in thin films is achieved via specular scattering. Our results provide a new paradigm for thermal manipulation since we show that smooth surfaces can create optimal reduction in thermal transport. Remarkably, we find that specular scattering not only reduces heat conduction but also provides the lowest thermal conductivity. The results lay the foundation for a broader and deeper understanding of nanoscale heat transport and the role of phonon surface scattering on thermal conductivity with applications in electronic, optoelectronic, and thermoelectric materials and devices.

In the past few decades, phonon transport in structures with length scales smaller than bulk phonon mean-free-paths has been studied theoretically by first-principle, molecular dynamics, and Boltzmann transport approaches [17-24]. For thin film phonon transport, it is found that diffuse phonon scattering at the boundaries causes a significant reduction of thermal conductivity with respect to bulk [25,26]. In order to show how the largest reduction in thin film thermal conductivity is achieved via specular scattering, we first need to discuss some basic concepts in thermal transport. Consider a temperature gradient $\nabla T$ in a bulk semiconductor [Fig. 1(a)]. Under the relaxation time approximation, phonon-phonon collisions maintain local thermodynamic equilibrium [1]. Phonons with wavevector **k** originating from collisions at point **x** contribute to the non-equilibrium thermal energy density $u$ at point $\mathbf{x}_0$ an amount given by the equilibrium energy density $u_{eq}$ at temperature $T(\mathbf{x})$, i.e. $u_{eq}(\mathbf{x})=u_{eq}[T(\mathbf{x})]$ times the probability $P$ for a phonon originating at point **x** to arrive at $\mathbf{x}_0$ after phonon-phonon collisions between **x** and $\mathbf{x}_0$ [27]. The probability of collision per unit length $P'$ is given by $(1/\ell_0)\exp(-r/\ell_0)$ where $\ell_0$ is the bulk phonon mean-free-path and $r$ the distance between **x** and $\mathbf{x}_0$ measured along the propagation direction given by the wavevector **k** [28]. Each phonon originating at point **x** contributes to the thermal current **J** at point $\mathbf{x}_0$ an amount equal to $\mathbf{J}=\mathbf{v}u$, where **v** is the phonon group velocity. The total thermal current **J** at $\mathbf{x}_0$ can be determined statistically by considering all thermal energy contributions from phonons originating at all points **x** in space. A net thermal current develops along the *x*-direction because, due to the thermal gradient, phonons originating on the left have larger energies than those on the right. Mathematically, for each wavevector **k**, the thermal current contribution $j_x$ at point $\mathbf{x}_0$ from phonons originating at point **x** is given by

$$j_x = v\cos\theta\, u_{eq}(x_0 - x)\, P(r) = v\cos\theta\, u_{eq}(x_0 - r\cos\theta)\, P(r) \tag{1}$$

where $v = |\mathbf{v}|$. To linear order in the temperature gradient, we have

$$j_x = -v\cos\theta\, \frac{\partial u_{eq}}{\partial x} r\cos\theta\, P(r) = -v\cos\theta\, \frac{\partial u_{eq}}{\partial T}\frac{\partial T}{\partial x} r\cos\theta\, P(r) \tag{2}$$

By considering all possible phonon origins $r$, the statistical analysis of phonon dynamics yields

$$\langle j_x \rangle_r = \int_0^\infty -v\cos\theta\, \frac{\partial u_{eq}}{\partial T}\frac{\partial T}{\partial x} r\cos\theta\, P'(r)\, dr = -v\cos\theta\, \frac{\partial u_{eq}}{\partial T}\ell_0 \cos\theta\, \frac{\partial T}{\partial x} \tag{3}$$

where $\int_0^\infty r P'(r)\, dr = \ell_0$. The total thermal current $J_x$ is obtained by considering all phonon wavevectors $\mathbf{k}$ and is given by

$$J_x = \langle\langle j_x \rangle_r\rangle_{\mathbf{k}} = -\frac{1}{(2\pi)^3}\int v\cos\theta\, \frac{\partial U_{eq}}{\partial T}\ell_0 \cos\theta\, d^3\mathbf{k}\, \frac{\partial T}{\partial x} \tag{4}$$

where $U_{eq}$ is the equilibrium thermal energy. Note that the thermal conductivity $\kappa$ can be obtained by using Fourier's law $J_x = -\kappa\, \partial T/\partial x$.

Next, let us consider a film with a temperature gradient along the in-plane direction [Fig. 1(b)]. In this case, when phonons originating at point **x** within the film are incident on the film boundaries at point **x'**, a proportion $p$ of these phonons is specularly reflected while the remaining $1-p$ proportion is diffusely scattered. Since diffuse surface scattering thermalizes phonons, a proportion $1-p$ of phonons thus originates at **x'**. Note that the phonon proportion $1-p$ originating at **x'** contributes an energy density $u(x_0-L_1\cos\theta)$ to the thermal current, which is smaller than the original energy density $u(x_0-r\cos\theta)$ at **x**. As a result, the thermal current in the film is reduced with respect to bulk. A statistical analysis of the thermal current accounting for all possible phonon origins **x** can be performed similarly to the bulk case. It requires statistical consideration of all multiple reflections and diffuse scattering events at the surfaces (see Tellier and Tossier [28] for details). For in-plane heat conduction, the statistical analysis yields

$$\int_0^\infty r P'(r) dr = \int_0^{L_1} \frac{r}{\ell_0} e^{-\frac{r}{\ell_0}} dr + (1-p)\int_{L_1}^\infty \frac{L_1}{\ell_0} e^{-\frac{r}{\ell_0}} dr + p\int_{L_1}^{L_1+L_2} \frac{r}{\ell_0} e^{-\frac{r}{\ell_0}} dr + ... \qquad (5)$$

and the thermal current can be calculated as [26]

$$J_x = -\frac{1}{(2\pi)^3}\int v\cos\theta \frac{\partial U_{eq}}{\partial T} \ell_0 \left[1 - \frac{(1-p)\exp(-L_1/\ell_0)}{1-p\exp(-L_2/\ell_0)}\right]\cos\theta d^3\mathbf{k}\frac{\partial T}{\partial x} \qquad (6)$$

The mean thermal current is obtained by integrating the thermal current $J_x$ over the thickness of the film from $z=0$ to $z=a$. The above statistical approach is equivalent to solving analytically the Boltzmann transport equation for a thin film as done by Fuchs and Sondheimer [29,30]. For in-plane heat conduction, completely diffuse surfaces ($p=0$) create the minimal thermal current while completely specular surfaces ($p=1$) have no effects on thermal transport and maintain the bulk thermal current [26].

We next consider thin films with a temperature gradient along the cross-plane direction [Fig. 1(c)]. We note that some basic principles that apply for bulk and in-plane heat conduction do not apply for cross-plane heat conduction. For example, for bulk and in-plane conduction there is a linear relation between the position $x$ along the $x$-axis and the position $r$ along the phonon path, that is $x=r\cos\theta$. This relationship is no longer valid for cross-plane transport [Fig. 1(c)]. Note that in cross-plane conduction, the value of $r$ is unlimited while the position $x$ has a maximum value given by the film thickness. This means that the thermal current, which arises from an statistical average of energies from phonons originating at different positions $x$, is no longer linearly correlated to the phonon mean-free-path $\ell$. In order to obtain the thermal current for cross-plane conduction, we correlate the position *r* in each phonon segment between successive reflections along the phonon path to the position *x* along the *x*-axis and perform the statistical analysis by considering all phonon origins and wavevectors. To linear order in the temperature gradient [31], the cross-plane thermal current contribution $j_x$ at point $\mathbf{x}_0$ from phonons originating at point **x** with wavevector **k**, is given by

$$j_x = v\cos\theta\, u_{eq}(x_0 - x)\, P(r) = -v\cos\theta \frac{\partial u_{eq}}{\partial x} x P(r) = -v\cos\theta \frac{\partial u_{eq}}{\partial T}\frac{\partial T}{\partial x} x P(r) \tag{7}$$

By considering all possible phonon origins, the statistical analysis yields

$$\langle j_x \rangle_r = \int_0^\infty -v\cos\theta \frac{\partial u_{eq}}{\partial T}\frac{\partial T}{\partial x} x P'(r)\, dr \tag{8}$$

where $\int_0^\infty x P'(r) dr = \left( \int_0^{L_1} \frac{r}{\ell_0} e^{-\frac{r}{\ell_0}} dr + (1-p)\int_{L_1}^\infty \frac{L_1}{\ell_0} e^{-\frac{r}{\ell_0}} dr + p\int_{L_1}^{2L_1} \frac{2L_1 - r}{\ell_0} e^{-\frac{r}{\ell_0}} dr + ...\right)\cos\theta \tag{9}$

The total thermal current $J_x$ is obtained by considering all phonon wavevectors **k** and is given by

$$J_x = -\frac{1}{(2\pi)^3}\int v\cos\theta \frac{\partial U_{eq}}{\partial T}\ell_0 \left[1 - \frac{(1+p)\exp(-L_1/\ell_0)}{1 + p\exp(-L_2/\ell_0)}\right]\cos\theta\, d^3\mathbf{k}\, \frac{\partial T}{\partial x} \tag{10}$$

The mean thermal current is calculated by integrating $J_x$ over the thickness of the film from $z=0$ to $z=a$ while considering thermal flux conservation. Note that Eq. 10 can alternatively be obtained by solving analytically the Boltzmann transport equation to linear order in temperature gradient [28,29] and introducing the antisymmetric characteristics of cross-plane thin film heat conduction.

Figure 2(a) shows our calculations for in-plane and cross-plane thermal conductivities for Si thin films at $T=300K$ for different thicknesses *a* and surface roughnesses $\eta$. Silicon bulk mean-free-paths and dispersion relations are taken from existing values in the literature [32,33]. Phonon-surface scattering effects are accounted for using the rigorous

Beckmann-Kirchhoff scattering theory extended with surface shadowing [34]. Note that in the past, Si cross-plane conductivities have only been reported for fully diffuse surfaces ($p=0$) [35-38]. Our calculations for $p=0$ show excellent agreement with literature. A remarkable result from Fig. 2 is the opposite dependence of thermal conductivity reduction with surface roughness. For in-plane heat conduction, κ is reduced with increasing diffuse surface scattering and the smallest value is obtained for fully diffusive surfaces ($p=0$). In contrast, for cross-plane heat conduction, κ is reduced with increasing specular surface scattering and the smallest value is obtained for fully specular surfaces ($p=1$). Note that while it is commonly accepted that increasing diffuse surface scattering provides the largest reduction in thermal conductivity, our results show that increasing specular surface scattering under cross-plane configuration is the mechanism to provide maximum reduction in thin film thermal conductivity.

In order to understand the observed behavior we next provide some insights for the underlying phonon dynamics. Let us first consider films with fully diffusive boundaries (i.e. $p=0$). In this case, for both in-plane and cross-plane heat conduction, we have $\int_0^\infty x P'(r) dr = \ell_0 (1 - e^{-\frac{L_1}{\ell_0}}) \cos\theta$. Note that phonons having small angles with respect to $\nabla T$ contribute more to the thermal current. In particular, when $\theta \to 0$, we have $L_1 \to \infty$ for in-plane conduction and $L_1 \to t/2$ for cross-plane conduction (Fig. 1). This significant difference in the value of $L_1$ for small angles explains why the cross-plane thermal conductivity is smaller than the in-plane thermal conductivity. In addition, to understand why the cross-plane conductivity is reduced with increasing specular scattering [Fig. 1(c)], we note that for increasing surface roughness a larger proportion of phonons are thermalized at the surfaces and, in particular, for $p=0$ all phonons reaching the surfaces are thermalized. This means that for $p=0$, the thermal current $j_x$ at point $\mathbf{x}_0$ is made of phonons originating at points within the film along $L_1$ (i.e. with no surface reflections) in addition to phonons originating at points $\mathbf{x}'$ at the surfaces. Phonons originating at $\mathbf{x}'$ on the left surface carry the largest energy values $u(x_0-L_1\cos\theta)$ while those originating at $\mathbf{x}'$ on the right surface carry the smallest energy values $u(x_0+L_1\cos\theta)$ due to the temperature gradient. The thermal current developed at $\mathbf{x}_0$ is thus maximal for $p=0$. Note that if we increase $p$, a proportion of phonons originating at the surfaces will be originating at points within the film and undergoing surface reflections before reaching $\mathbf{x}_0$. Phonons originating within the film have intermediate energy values $u(x_0-L_1\cos\theta)<u<u(x_0+L_1\cos\theta)$ and do not provide the maximal energy exchange because their difference with respect to $u(x_0)$ is reduced. This is opposite to in-plane conduction [Fig. 1(b)] where increasing the surface specularity $p$ translates into phonons with larger energies than $u(x_0-L_1\cos\theta)$ on the left and smaller energies than $u(x_0+L_1\cos\theta)$ on the right, thus increasing the difference with respect to $u(x_0)$. The smaller phonon energy differences from $u(x_0)$ for increasing $p$ explain why the cross-plane thermal conductivity is reduced with increasing specular scattering.

Another important aspect of nanoscale heat conduction is to establish the amount of heat carried by phonons with different wavelengths. Such heat wavelength spectrum can be used to rationally design thermal materials [39]. Figure 3(a) shows the heat wavelength spectrum for in-plane and cross-plane heat transport for a 10nm-thick Si thin film for a wide range of surface conditions $p=0$, $\eta=1.0, 0.5, 0.3, 0.2, 0.1$nm and $p=1$ at $T=300$K. For in-plane conduction, the spectrum shifts to short wavelengths when diffuse surface scattering is increased. This is a consequence of the reduction in the phonon mean-free-paths of long wavelength phonons [26]. However, for cross-plane conduction this feature is reversed and the spectrum shifts to short wavelengths when specular surface scattering is increased. This is consistent and explained by the following consideration. For $p=0$, a large portion of long wavelength phonons originate at the surfaces due to their large phonon-phonon mean fee paths $\ell_0$ (and surface thermalization). As $p$ is increased, however, these phonons originate within the film and undergo surface reflections before reaching $\mathbf{x}_0$. As previously discussed, the energy contribution of these phonons to the thermal current is reduced. As a result, the wavelength spectrum shifts to short wavelengths with increasing $p$, which is opposite to in-plane heat conduction.

Given the fundamental differences between in-plane and cross-plane conduction, a question that arises is how different are the thermal conductivities along the in-plane and cross-plane directions, or equivalently, what is the degree of anisotropy in the film. We calculate in Figure 4 the ratio $\kappa_{in}/\kappa_{cross}$ between in-plane and cross-plane conductivities as a function of thickness $a$ for different surface roughness $\eta$ ranging from fully specular to fully diffuse surfaces. Note that thin films with fully diffuse surfaces ($p=0$) provide the smallest thermal anisotropy, with a maximum ratio $\kappa_{in}/\kappa_{cross}$ ~4. As the specularity of the surface increases (i.e. smaller roughness $\eta$), the ratio $\kappa_{in}/\kappa_{cross}$ increases significantly and heat conduction becomes more anisotropic. In particular, for fully specular surfaces ($p=1$), the largest anisotropic behavior is observed and in-plane heat conduction is highly preferred over cross-plane conduction.

It is also important to note that for some phonon frequencies, multiple reflections of phonons at the film surfaces can give rise to coherent interference. The role of coherent interference is to create guided phonon modes along the plane of the film [40]. For in-plane conduction, these modes have reduced group velocities in the direction of the thermal gradient [41]. However, we highlight here that for cross-plane conduction these phonon modes have *zero* group velocities in the direction of the thermal gradient. That is, guided modes transport no energy across the film. We calculate the proportion of the thermal conductivity that can be subject to coherent interference for cross-plane conduction by analyzing the phonon wavelength spectrum and the number of multiple reflections [34]. We found that for $t > 10$nm, the proportion of thermal conductivity reduction due to phonons with zero group velocities along the thermal gradient is less than 1% regardless the surface roughness. However for $t=4$nm, these proportions are 12%, 7.6%, and 1.5% for $p=1$, $\eta=0.1$nm, and $\eta=0.2$nm, respectively. We note that the amount of thermal phonons with zero group velocities can be increased further by considering other materials (e.g. SiGe) or lower temperatures [42].

In conclusion, we have shown that specular surface scattering provides the largest reduction in cross-plane thin film thermal conductivity. This is in contrast to current understanding on thermal phonon transport in thin films where diffuse surface scattering is used to reduce the thermal conductivity. Our results create a new paradigm to manipulate heat conduction since smooth surfaces – in contrast to rough surfaces – are more effective on suppressing heat conduction. The fundamental understanding introduced here can have a strong impact on the way we use diffuse and specular scattering to manipulate thermal conductivities in energy, electronic, and optoelectronic materials and devices.

**Figure Legends**

FIG. 1 Schematics for thermal phonon transport in (a) bulk, (b) in-plane, and (c) cross-plane thin-film heat conduction. Phonons with wavevector **k** originate at **x** with energy density $u$ and contribute to the thermal current at point $\mathbf{x_0}$ after traveling a path $r$ under phonon-phonon collisions. Red areas at points **x'** represent the diffuse scattering of phonons at the surfaces.

FIG. 2 Thermal conductivity κ as a function film thickness $a$ for in-plane (top) and cross-plane (bottom) silicon heat conduction for different surface roughnesses $\eta$ and fully diffuse ($p=0$) and fully specular ($p=1$) surfaces.

FIG. 3 Heat wavelength spectrum for a 10nm-thick Si film for (a) in-plane and (b) cross-plane heat conduction.

FIG. 4 Anisotropic thin film properties ($\kappa_{IN}/\kappa_{CROSS}$) as a function of film thickness and for different surface conditions showing maximum anisotropy for specular surfaces.

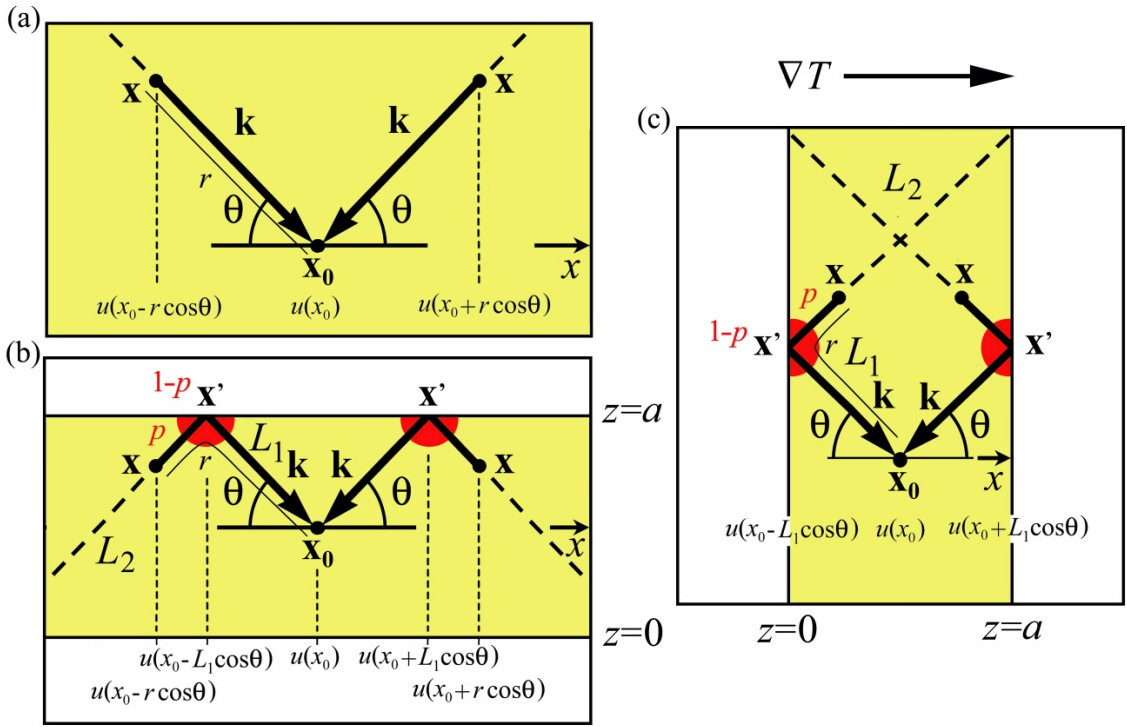

**Figure 1**

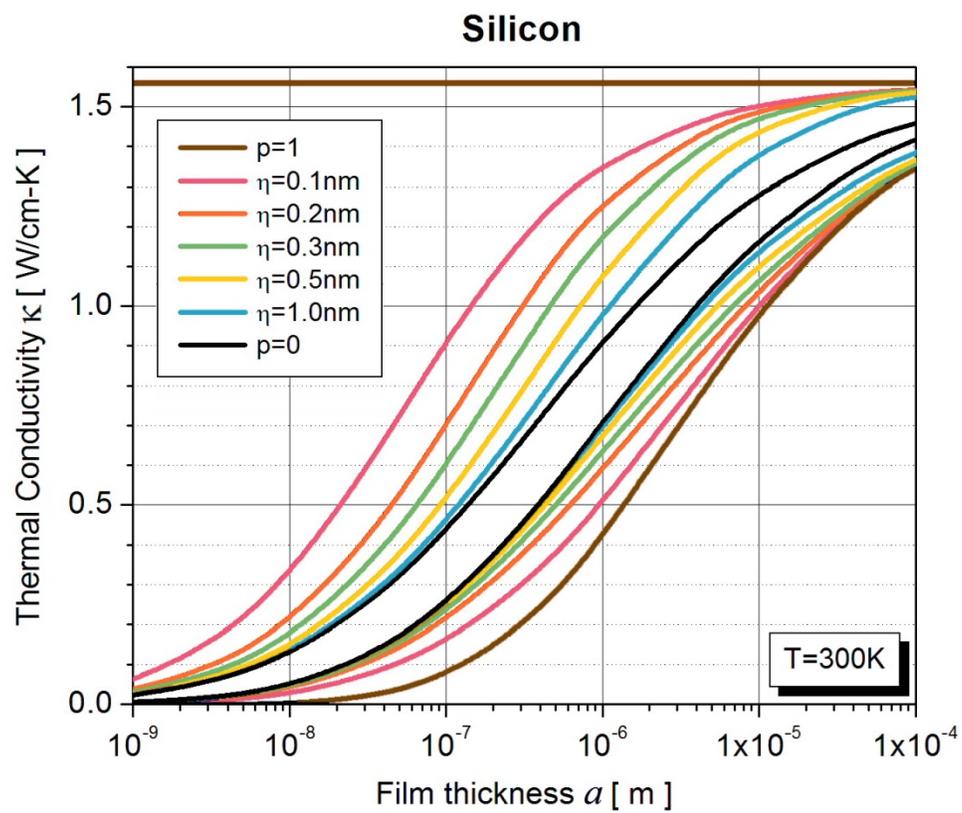

**Figure 2**

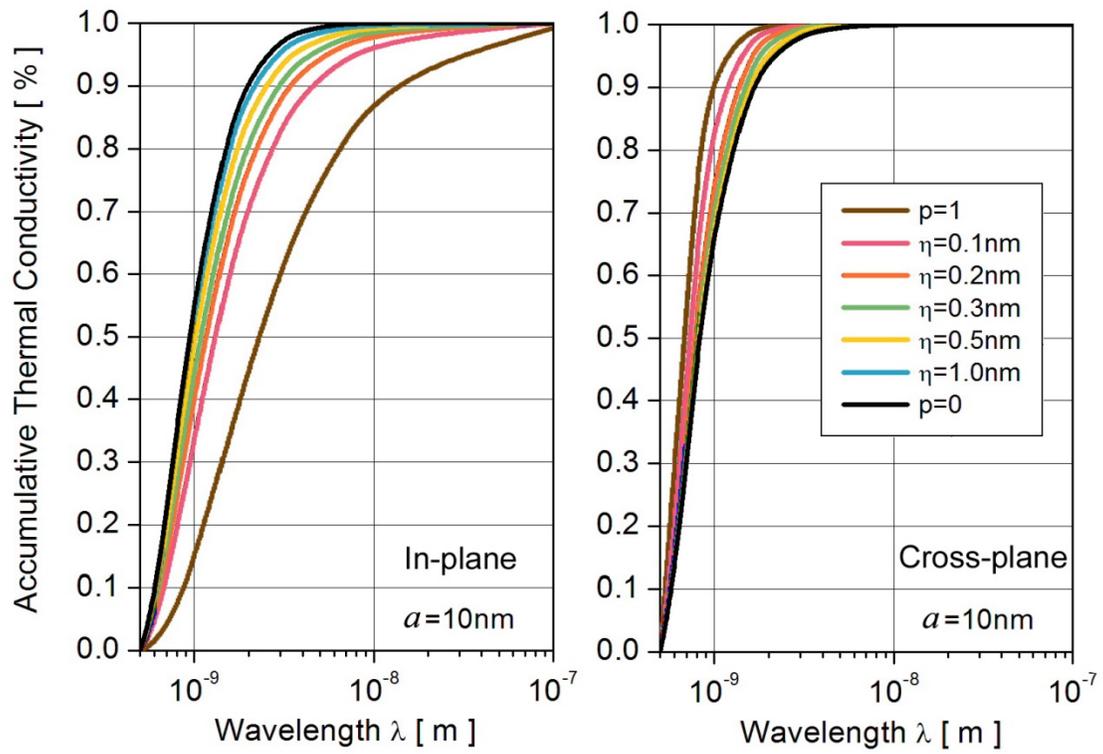

**Figure 3**

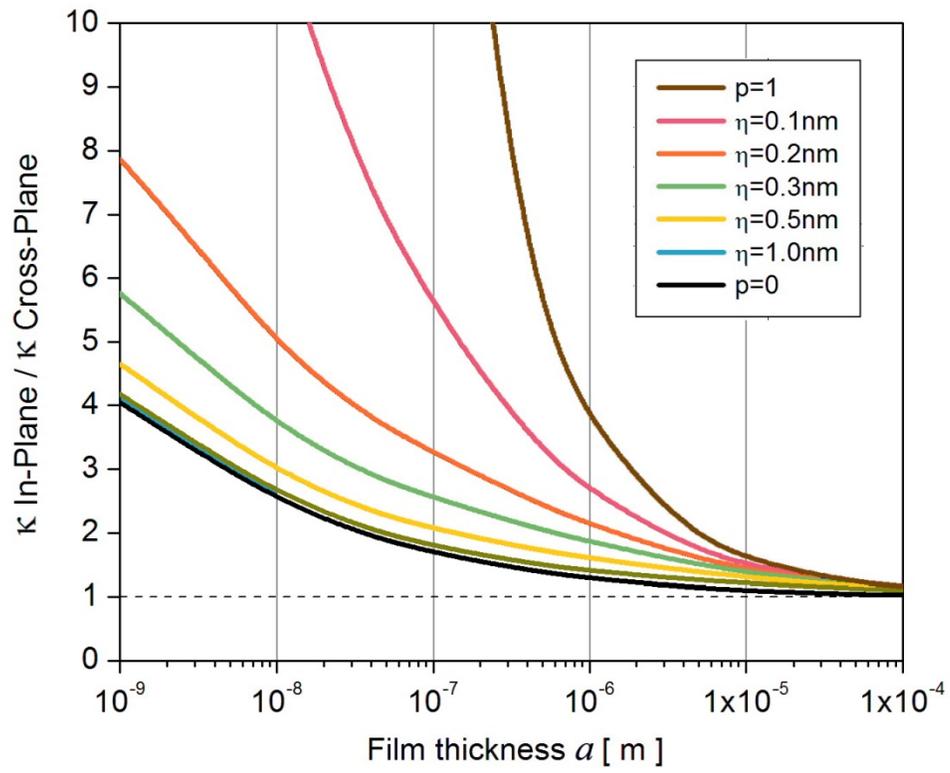

**Figure 4**